\documentclass[aps,preprint]{revtex4}%
\usepackage{amsfonts}
\usepackage{amsmath}
\usepackage{amssymb}
\usepackage{graphicx}%
\providecommand{\U}[1]{\protect\rule{.1in}{.1in}}
\begin{document}
\author{}
\title{Revisiting the tunneling spectrum and information recovery of a general charged and rotating black hole}
\author{Ge-Rui Chen}
\email{chengerui@emails.bjut.edu.cn}
\author{Yong-Chang Huang}
%\email{ ychuang@bjut.edu.cn}
\affiliation{Institute of Theoretical Physics, Beijing University of
Technology, Beijing, 100124, China}

\begin{abstract}
In this paper we revisit the tunneling spectrum of a charged and
rotating black hole--Kerr-Newman black hole by using Parikh and
Wilczek' s tunneling method and get the most general result compared with the
works \cite{jwc,jz}. We find an ambiguity in Parikh and Wilczek' s tunneling method, and give a reasonable description.
We use this general spectrum to discuss the information recovery based on the Refs. \cite{bcz1,bcz2,bcz3}.
For the tunneling spectrum we
obtained, there exit correlations between sequential Hawking radiations, information can be carried out by
such correlations, and the entropy is conserved during the radiation
process. So we resolve the information loss paradox based on the methods \cite{bcz1,bcz2,bcz3} in 
the most general case.\\
PACS: {04.70.Dy; 04.70.-s; 03.65.Sq; 03.67.-a}\\
Keywords: {\ Kerr-Newman black hole, tunneling spectrum,
correlation, mutual information, entropy conservation process}
\end{abstract}

\maketitle

\section{\textbf{Introduction}}
Since Hawking radiation \cite{sk1,sk2} was discovered, black hole
physics has been greatly developed. People know that black hole is
not the object from which gravity prevents anything, including
light, from escaping, and it can radiate thermal spectrum. Black
holes are not the final state of stars, and, with the emission of
Hawking radiation, they lose energy, shrink, and eventually
evaporate completely. However, because the radiation is a purely
thermal spectrum, it also sets up a disturbing and difficult
information loss problem: what happens to information during black
hole evaporation? This scenario is inconsistent with the unitary
principle of quantum mechanics \cite{swk3,swk4,cgc,sdm}.

The semi-classical method which models Hawking radiation as a
tunneling process was proposed by Parikh and Wilczek
\cite{mkp1,mkp3} in 2001. This method uses WKB approximation, and
calculates the imaginary part of the action $Im
I=\int_{r_{in}}^{r_{out}}p_r dr$ for the classically forbidden
process across the horizon. The tunneling probability is given by
$\Gamma\propto\exp(-2Im I)$. They found that considering the back
reaction of the emission particle to the spacetime, a non-thermal
spectrum is given, which supports the underlying unitary theory.
After that, there were many works to generalize this method to other
different kinds of black holes and all got the same results. The
most general case is the charged particles' tunneling from a charged
and rotating black hole--Kerr-Newman black hole \cite{jwc,jz}.

In 2009, Refs. \cite{bcz1,bcz2,bcz3} gave more detailed discussions
about Parikh and Wilczek's non-thermal spectrum. Their results are
that there exist correlations among sequential Hawking radiations, the
correlation is equal to mutual information, and black hole radiation
is an entropy conservation process. In Ref. \cite{bcz2}, they
discussed the properties of charged particles' tunneling spectrum
from Kerr-Newman black hole by using the tunneling spectrum in Refs.
\cite{jwc,jz}. We find that this spectrum is not the most general
case, and is just the special case for $j=a\omega$ (see the end of section 2). We revisit the calculation and get the most
general tunneling spectrum (\ref{c4}). We also find an ambiguity in Parikh and Wilczek's method, and give a reasonable description. Then we use the tunneling spectrum obtained to
discuss information conservation and generalize Refs.
\cite{bcz1,bcz2,bcz3}' results to the most general case.

This paper is organized as follows. In Section 2, we revisit the
tunneling method to get a general non-thermal spectrum from
Kerr-Newman black hole. In Section 3, we investigate the properties
of this non-thermal spectrum. In the last Section, we give some
discussions and conclusions.

\section{Revisiting Parikh and Wilczek' s tunneling method from Kerr-Newman black hole}
We recalculate the tunneling spectrum from a general charged and
rotating black hole--Kerr-Newman black hole. Although there are some
works \cite{jwc,jz} about this problem, our discussion has new
insights and gets the most general result. We will use directly
some results already in Ref. \cite{jz}.

According to WKB approximation, the emission rate $\Gamma$ can be
given as
\begin{eqnarray}
\Gamma\sim\exp(-2Im I),
\end{eqnarray}
where $I$ is the action of the emitting particle. In the
Painlev\'{e}-Kerr-Newnan coordinate system, the action for the
classically forbidden trajectory of an emitting particle is
\begin{eqnarray}
I&=&\int[p_{r}dr-p_{A_t}dA_{t}-p_{\phi}d\phi]\nonumber\\
 &=&\int
 [p_{r}dr-\frac{p_{A_t}\dot{A_t}}{\dot{r}}dr-\frac{p_{\phi}\dot{\phi}}{\dot{r}}dr]\nonumber\\
 &=&\int\int[dp_r-\frac{\dot{A_t}dp_{A_t}}{\dot{r}}-\frac{\dot{\phi}dp_{\phi}}{\dot{r}}]dr.
\end{eqnarray}
For the emitting particle, the Hamilton's equations are
\begin{eqnarray}
\dot{r}&=&\frac{dH}{dp_r}|_{(r;A_t,p_{A_t};\phi,p_\phi)}=\frac{d\omega}{dp_r},\nonumber\\
\dot{A_t}&=&\frac{dH}{dp_{A_t}}|_{(A_t;r,p_{r};\phi,p_\phi)}=\frac{V_0dq}{dp_{A_t}},
\end{eqnarray}
so we have
\begin{eqnarray}
dp_r=\frac{d\omega}{\dot{r}},\ \dot{A_t}dp_{A_t}=V_0dq,
\end{eqnarray}
where $\omega,q$ are the energy and charge of the emitting particle.
$V_0$ is the electrostatic potential on the outer event horizon
$r=r_+$,
\begin{eqnarray}
V_0=\frac{Qr_+}{r_+^2+a^2},
\end{eqnarray}
where $Q$ and $a=\frac{J}{M}$ are the charge and angular momentum of
unit mass of black hole respectively. So we can get
\begin{eqnarray}
ImI&=&Im\int\int[\frac{d\omega}{\dot{r}}-\Omega\frac{dj}{\dot{r}}-\frac{V_0dq}{\dot{r}}]dr\nonumber\\
&=&\int\int\frac{dr}{\dot{r}}[d\omega-\Omega dj-V_0dq],\label{c2}
\end{eqnarray}
where $\Omega=\frac{a}{r_+^2+a^2}$ is the angular velocity of the
horizon and $p_{\varphi}=j$ is the angular momentum of the emitting
particle. We consider the s-wave, so $\dot{\phi}=\Omega$.

In Ref \cite{jz}. the expression of $\dot{r}$  for the charged
particle is
\begin{eqnarray}
\dot{r}=\frac{\Delta}{2}\sqrt{\frac{\rho^2}{(\rho^2-\Delta)[(r^2+a^2)^2-\Delta
a^2\sin^2\theta]}},\label{c1}
\end{eqnarray}
where
\begin{eqnarray}
\Delta&=&r^2+a^2+Q^2-2Mr=(r-r_+)(r-r_-),\nonumber\\
r_{\pm}&=&M\pm\sqrt{M^2-a^2-Q^2},\nonumber\\
\rho^2&=&r^2+a^2\cos^2\theta.
\end{eqnarray}
Putting Eq. (\ref{c1}) into Eq. (\ref{c2}), we have
\begin{eqnarray}
ImI&=&Im\int\int\frac{2\sqrt{(\rho^2-\Delta)[(r^2+a^2)^2-\Delta
a^2\sin^2\theta]}}{(r-r_+)(r-r_-)\sqrt{\rho^2}}dr(d\omega-\Omega
dj-V_0dq)\nonumber\\
&=&Im\int \pi i 2\frac{r_+^2+a^2}{r_+-r_-}(d\omega-\Omega
dj-V_0dq)\nonumber\\
&=&2\pi\int\frac{r_+^2+a^2}{r_+-r_-}(d\omega-\Omega dj-V_0dq)\nonumber\\
&=&2\pi\int\frac{r_+^2+a^2}{r_+-r_-}d\omega-\frac{a}{r_+-r_-}dj-\frac{Qr_+}{r_+-r_-}dq,\label{c3}
\end{eqnarray}
where the integral of $r$ is done by deforming the contour around
the pole in the second equality.

When particle's self-gravitation is taken into account we should
replace $M,J,Q$ with $M-\omega, J-j, Q-q$ in Eq. (\ref{c3}),
\begin{eqnarray}
ImI&=&\pi\int_{(0,0,0)}^{(\omega,j,q)}[2(M-\omega)+\frac{2(M-\omega)^3-(Q-q)^2(M-\omega)}{\sqrt{(M-\omega)^4-(J-j)^2-(Q-q)^2(M-\omega)^2}}]d\omega\nonumber\\
&&-\frac{J-j}{\sqrt{(M-\omega)^4-(J-j)^2-(Q-q)^2(M-\omega)^2}}dj\nonumber\\
&&-[(Q-q)+\frac{(Q-q)(M-\omega)^2}{\sqrt{(M-\omega)^4-(J-j)^2-(Q-q)^2(M-\omega)^2}}]dq.
\end{eqnarray}
It is easy to find that
\begin{eqnarray}
ImI&=&-\frac{1}{2}\int[\frac{\partial(\Delta S)}{\partial
\omega}d\omega+\frac{\partial(\Delta S)}{\partial
j}dj+\frac{\partial(\Delta S)}{\partial q}dq]\nonumber\\
&=&-\frac{1}{2}\Delta S,
\end{eqnarray}
where
\begin{eqnarray}
\Delta S&=&\pi[2(M-\omega)^2-(Q-q)^2+2\sqrt{(M-\omega)^4-(J-j)^2-(Q-q)^2(M-\omega)^2}]\nonumber\\
&&-\pi[2M^2-Q^2+2\sqrt{M^4-J^2-Q^2M^2}]\label{h2}
\end{eqnarray}
is the change of black hole entropy after emitting a particle with
$(\omega,j,q)$. So the tunneling rate is
\begin{eqnarray}
\Gamma&\sim&\exp(-2ImI)=\exp(\Delta S)\nonumber\\
&=&\exp(\pi[2(M-\omega)^2-(Q-q)^2+2\sqrt{(M-\omega)^4-(J-j)^2-(Q-q)^2(M-\omega)^2}]\nonumber\\
&&-\pi[2M^2-Q^2+2\sqrt{M^4-J^2-Q^2M^2}]).\label{c4}
\end{eqnarray}
This is the general result for the tunneling spectrum. Refs.
\cite{jwc,jz} only give a special case with $j=a\omega$,
\begin{eqnarray}
\Gamma&\sim&\exp(-2ImI)\nonumber\\
&=&\exp(\pi[2(M-\omega)^2-(Q-q)^2+2(M-\omega)\sqrt{(M-\omega)^2-a^2-(Q-q)^2}]\nonumber\\
&&-\pi[2M^2-Q^2+2M\sqrt{M^2-a^2-Q^2}]).
\end{eqnarray}

Let us give some discussions. In the original paper of Parikh and Wilczek's tunneling method \cite{mkp1,mkp3} and many other followed works \cite{jwc,jz}, for the Hamilton's equation $\dot{r}=\frac{dH}{dp_r}$, they argued that $H=M-\omega$ represents the Hamiltonian of the black hole, and
$\dot{r}$ is the radial path of the emitting particles. We think
there exists an ambiguity since we could not tell the Hamilton's equation is for what physical object. In Ref. \cite{mkp1,mkp3,jwc,jz}, a minus sign appears in equation $dH=-d\omega$, and integration around the pole gives the result $-\pi i Res $, where the minus sign comes from the assumption $r_{in}>r_{out}$. This integration is inconsistent with that of another tunneling method--complex path integration method (also called  Hamilton-Jacobi method) which gives the integration $ \pi i Res $ \cite{tp,rk2,rk3}. In our article, we consider that the Hamilton's equation is for the emitting particles, and $H=\omega$ represents the energy of emitting particles. The integrating around pole is $ \pi i Res $ which is the same as that of the complex path integration method. In our opinion the two different tunneling methods should have the same integration around the pole since this integration reflects the barrier in the tunneling process.

\section{Information recovery of tunneling spectrum from Kerr-Newman black hole}
In this section we investigate the properties of the tunneling
spectrum (\ref{c4}) based on Refs. \cite{bcz1,bcz2,bcz3}'
methods. From Eq. (\ref{c4}), the probability for the emission of a particle with energy, angular momentum and charge ($\omega_1, j_1,q_1$) is
\begin{eqnarray}
\Gamma(\omega_1,j_1,q_1)&=&\exp(\pi[2(M-\omega_1)^2-(Q-q_1)^2+2\sqrt{(M-\omega_1)^4-(J-j_1)^2-(Q-q_1)^2(M-\omega_1)^2}]\nonumber\\
&&-\pi[2M^2-Q^2+2\sqrt{M^4-J^2-Q^2M^2}]).
\end{eqnarray}
And the probability for the emission of a particle with energy, angular momentum and charge ($\omega_2, j_2,q_2$) is
\begin{eqnarray}
\Gamma(\omega_2,j_2,q_2)&=&\exp(\pi[2(M-\omega_2)^2-(Q-q_2)^2+2\sqrt{(M-\omega_2)^4-(J-j_2)^2-(Q-q_2)^2(M-\omega_2)^2}]\nonumber\\
&&-\pi[2M^2-Q^2+2\sqrt{M^4-J^2-Q^2M^2}]).\label{P3}
\end{eqnarray}
Note that ($\omega_1,j_1,q_1$) and ($\omega_2,j_2,q_2$) represent two independent emitting particles, so the expressions
should have the same form.

Let us consider a process as follows. One particle with energy, angular
momentum and charge ($\omega_1,j_1,q_1$) emits, then another
particle with energy, angular momentum and charge ($\omega_2,j_2,q_2$) emits. The probability for the
emission of the second particle is
\begin{eqnarray}
\Gamma(\omega_2,j_2,q_2|\omega_1,j_1,q_1)&=&\exp(\pi[2(M-\omega_1-\omega_2)^2-(Q-q_1-q_2)^2\nonumber\\
&+&2\sqrt{(M-\omega_1-\omega_2)^4-(J-j_1-j_2)^2-(Q-q_1-q_2)^2(M-\omega_1-\omega_2)^2}]\nonumber\\
&-&\pi[2(M-\omega_1)^2-(Q-q_1)^2\nonumber\\
&+&2\sqrt{(M-\omega_1)^4-(J-j_1)^2-(Q-q_1)^2(M-\omega_1)^2}]),\label{h1}
\end{eqnarray}
which is the conditional probability and is different from the
independent probability (\ref{P3}). So the emitting probability for
the two successive emissions can be calculated as
\begin{eqnarray}
\Gamma(\omega_1,j_1,q_1,\omega_2,j_2,q_2)&\equiv&\Gamma(\omega_1,j_1,q_1)\Gamma(\omega_2,j_2,q_2|\omega_1,j_1,q_1)\nonumber\\
&=&\exp(\pi[2(M-\omega_1)^2-(Q-q_1)^2\nonumber\\
&&+2\sqrt{(M-\omega_1)^4-(J-j_1)^2-(Q-q_1)^2(M-\omega_1)^2}]\nonumber\\
&&-\pi[2M^2-Q^2+2\sqrt{M^4-J^2-Q^2M^2}])\nonumber\\
&&\times\exp(\pi[2(M-\omega_1-\omega_2)^2-(Q-q_1-q_2)^2\nonumber\\
&&+2\sqrt{(M-\omega_1-\omega_2)^4-(J-j_1-j_2)^2-(Q-q_1-q_2)^2(M-\omega_1-\omega_2)^2}]\nonumber\\
&&-\pi[2(M-\omega_1)^2-(Q-q_1)^2\nonumber\\
&&+2\sqrt{(M-\omega_1)^4-(J-j_1)^2-(Q-q_1)^2(M-\omega_1)^2}])\nonumber\\
&=&\exp(\pi[2(M-\omega_1-\omega_2)^2-(Q-q_1-q_2)^2\nonumber\\
&&+2\sqrt{(M-\omega_1-\omega_2)^4-(J-j_1-j_2)^2-(Q-q_1-q_2)^2(M-\omega_1-\omega_2)^2}]\nonumber\\
&&-\pi[2M^2-Q^2+2\sqrt{M^4-J^2-Q^2M^2}]).
\end{eqnarray}
The last equality is
$\Gamma(\omega_{1}+\omega_{2},j_{1}+j_{2},q_1+q_2)$, so we have
\begin{eqnarray}
\Gamma(\omega_1,j_1,q_1,\omega_2,j_2,q_2)=\Gamma(\omega_{1}+\omega_{2},j_{1}+j_{2},q_1+q_2).
\end{eqnarray}
This is an important relationship which tells us that the
probability for two particles emitting successively with energy, angular momenta and charges ($\omega_1,j_1,q_1$) and ($\omega_2,j_2,q_2$) is
the same as the probability for one particle emitting with energy, angular
momentum and charge $(\omega_{1}+\omega_{2},j_{1}+j_{2},q_1+q_2)$. It is easy to see that
\begin{eqnarray}
\Gamma(\omega_1,j_1,q_1,\omega_2,j_2,q_2,\cdots,\omega_i,j_i,q_i)
&=&\Gamma(\omega_1,j_1,q_1)\Gamma(\omega_2,j_2,q_2|\omega_1,j_1,q_1)\nonumber\\
&&\times\cdots\times
\Gamma(\omega_i,j_i,q_i|\omega_1,j_1,q_1,\cdots,\omega_{i-1},j_{i-1},q_{i-1})\nonumber\\
&=&\Gamma(\omega_1+\cdots+\omega_i,j_1+\cdots+j_i,q_1+\cdots+q_i).\label{P5}
\end{eqnarray}
This is an important relationship we will use later.

For the Hawking radiation, the correlation between the two
sequential emissions can be calculated as \cite{bcz1,bcz2,bcz3}
\begin{eqnarray}
\ln\Gamma(\omega_1+\omega_2,j_1+j_2,q_1+q_2)-\ln[\Gamma(\omega_1,j_1,q_1)\Gamma(\omega_2,j_2,q_2)]
&=&\ln\frac{\Gamma(\omega
_1+\omega_2,j_1+j_2,q_1+q_2)}{\Gamma(\omega_1,j_1,q_1)\Gamma(\omega_2,j_2,q_2)}\nonumber\\
&=&\ln\frac{\Gamma(\omega_1,j_1,q_1)\Gamma(\omega_2,j_2,q_2|\omega_1,j_1,q_1)}{\Gamma(\omega_1,j_1,q_1)\Gamma(\omega_2,j_2,q_2)}\nonumber\\
&=&\ln\frac{\Gamma(\omega_2,j_2,q_2|\omega_1,j_1,q_1)}{\Gamma(\omega_2,j_2,q_2)}\neq0,
\end{eqnarray}
which shows that the two emissions are statistically dependent, that
is to say, there are correlations between sequential Hawking
radiations.

Like Eq. (\ref{h1}) the conditional probability
$\Gamma(\omega_i,j_i,q_i|\omega_1,j_1,q_1,\cdots,\omega_{i-1},j_{i-1},q_{i-1})$
is the tunneling probability of a particle emitting with energy,
angular momentum and charge ($\omega_i,j_i,q_i$) after a sequence of
radiation from $1 \rightarrow (i-1)$, so conditional entropy taken
away by this tunneling particle is given by
\begin{eqnarray}
S(\omega_i,j_i,q_i|\omega_1,j_1,q_1,\cdots,\omega_{i-1},j_{i-1},q_{i-1})&=&-\ln\Gamma(\omega_i,j_i,q_i|\omega_1,j_1,q_1,\cdots,\omega_{i-1},j_{i-1},q_{i-1}).\nonumber\\
\end{eqnarray}

The mutual information for the emissions of two particles with
energy, angular momenta and charges $(\omega_1,j_1,q_1)$ and
$(\omega_2,j_2,q_2)$ is defined as \cite{bcz1,bcz2,bcz3}
\begin{eqnarray}
S(\omega_2,j_2,q_2:\omega_1,j_1,q_1)&\equiv&
S(\omega_2,j_2,q_2)-S(\omega_2,j_2,q_2|\omega_1,j_1,q_1)\nonumber\\
&=&-\ln\Gamma(\omega_2,j_2,q_2)+\ln\Gamma(\omega_2,j_2,q_2|\omega_1,j_1,q_1)\nonumber\\
&=&\ln\frac{\Gamma(\omega_2,j_2,q_2|\omega_1,j_1,q_1)}{\Gamma(\omega_2,j_2,q_2)},
\end{eqnarray}
which shows that mutual information is equal to the correlation
between the sequential emissions, that is to say, the information is
encoded in the correlations between Hawking radiations.

Let us calculate the entropy carried out by Hawking radiations. The
entropy carried out by the first emitting particle with energy, angular
momentum and charge $(\omega_1,j_1,q_1)$ is
\begin{eqnarray}
S(\omega_1,j_1,q_1)&=&-\ln\Gamma(\omega_1,j_1,q_1).
\end{eqnarray}
The conditional entropy carried out by the second emitting particle after the first
emission is
\begin{eqnarray}
S(\omega_2,j_2,q_2|\omega_1,j_1,q_1)=-\ln\Gamma(\omega_2,j_2,q_2|\omega_1,j_1,q_1).
\end{eqnarray}
So the total entropy carried out by the two sequential emissions is
\begin{eqnarray}
S(\omega_1,j_1,q_1,\omega_2,j_2,q_2)=S(\omega_1,j_1,q_1)+S(\omega_2,j_2,q_2|\omega_1,j_1,q_1).
\end{eqnarray}

Assuming the black hole exhausts after radiating $n$ particles, we
have the following relationship
\begin{eqnarray}
\sum_i^n\omega_i=M, \sum_i^nj_i=J, \sum_i^nq_i=Q,
\end{eqnarray}
where $M,J,Q$ are the mass, angular momentum and charge of the black
hole. The entropy carried out by all the emitting particles is
\begin{eqnarray}
S(\omega_1,j_1,q_1,\cdots,\omega_{n},j_{n},q_n)&=&\sum_{i=1}^{n}S(\omega_i,j_i,q_i|\omega_1,j_1,q_1,\cdots,\omega_{i-1},j_{i-1},q_{i-1})\nonumber\\
&=&S(\omega_1,j_1,q_1)+S(\omega_2,j_2,q_2|\omega_1,j_1,q_1)\nonumber\\
&&+\cdots+S(\omega_n,j_n,q_n|\omega_1,j_1,q_1,\cdots,\omega_{n-1},j_{n-1},q_{n-1})\nonumber\\
&=&-\ln\Gamma(\omega_1,j_1,q_1)-\ln\Gamma(\omega_2,j_2,q_2|\omega_1,j_1,q_1)-\cdots\nonumber\\
&&-\ln\Gamma(\omega_n,j_n,q_n|\omega_1,j_1,q_1,\cdots,\omega_{n-1},j_{n-1},q_{n-1})\nonumber\\
&=&-\ln[\Gamma(\omega_1,j_1,n_1)\times\Gamma(\omega_2,j_2,q_2|\omega_1,j_1,q_1)
\times\cdots\times\nonumber\\
&&\Gamma(\omega_n,j_n,q_n|\omega_1,j_1,q_1,\cdots,\omega_{n-1},j_{n-1},q_{n-1})]\nonumber\\
&=&-\ln\Gamma(\omega_1+\omega_2+\cdots+\omega_n,j_1+j_2+\cdots+j_n,q_1+q_2+\cdots+q_n)\nonumber\\
&=&-\ln\Gamma(M,J,Q)=\pi[2M^2-Q^2+2\sqrt{M^2-J^2-Q^2M^2}]\nonumber\\
&=&S_{BH},
\end{eqnarray}
where we use the Eq. (\ref{P5}) in the fifth equation and Eq. (\ref{h2}) in the last equation. The result
shows that the entropy carried out by all the emitting particles equals to the black hole original entropy, so the total entropy is
conserved in the process of radiation.

\section{\textbf{Discussions and Conclusions}}
In this paper we recalculate the tunneling spectrum of a general
charged and rotating black hole--Kerr-Newman black hole by using
Parikh and Wilczek' s method and get the most general result
(\ref{c4}). A technical difference between our work and former works
\cite{jwc,jz} is that we only use the result $\dot{\phi}=\Omega$
since for the s-wave the particles radiate along the normal
direction of cross section, and the condition $j=a\omega$ used in
Refs. \cite{jwc,jz} is unnecessary. We also discuss the ambiguity in Parikh and Wilczek' s tunneling method, and compare it with another tunneling method--complex path integration method (also called  Hamilton-Jacobi method). We use this most general result
to investigate the back hole information recovery based on Refs. \cite{bcz1,bcz2,bcz3}, and find that for this
tunneling spectrum there exit correlations between sequential Hawking radiations, information can be
carried out by such correlations, and the entropy is conserved
during the radiation process. Therefore, the most general case about
back hole information recovery based on the methods \cite{bcz1,bcz2,bcz3} is given in
this paper.

\begin{acknowledgements}
This work is supported by National Natural Science Foundation of
China (No.11275017 and No.11173028).
\end{acknowledgements}

\end{document}